\def\nc{\newcommand}
\nc{\be}{\begin{equation}}
\nc{\ee}{\end{equation}}
\nc{\bea}{\begin{eqnarray}}
\nc{\eea}{\end{eqnarray}}
\def\nn{\nonumber}

\nc{\el}{\text{\boldmath $l$}}
\nc{\en}{\text{\boldmath $n$}}
\nc{\emm}{\text{\boldmath $m$}}
\nc{\emmB}{\text{\boldmath $\overline{m}$}}
\nc{\emP}{\emm+\emmB}
\nc{\emM}{\emm-\emmB}
\nc{\deltaB}{\overline{\delta}}
\nc{\cQ}{\mathcal{Q}}
\nc{\xd}{\mathbf {d}}
\nc{\rt}{\sqrt{2}}
\nc{\pd}{\partial}
\nc{\ddx}[1]{\frac{\pd}{\pd #1}}
\nc{\dx}[1]{\xd #1}
\nc{\pref}[1]{(\ref{#1})}

\documentclass[aps,prd,preprint,double-spaced,nofootinbib,showpacs]{revtex4}

\begin{document}

\title{On a choice of the Bondi radial coordinate 
and news function \\ for the axisymmetric two-body problem.}

\author{S. Bonanos}
\email[]{sbonano@inp.demokritos.gr}

\affiliation{Institute of Nuclear Physics, NCSR Demokritos,\\
15310 Aghia Paraskevi, Attiki, Greece}

\date{\today}

\begin{abstract}
    In the Bondi formulation of the axisymmetric vacuum Einstein equations,
    we argue that the ``surface area'' coordinate condition  determining the
    ``radial'' coordinate can be considered as part of the initial data 
    and should  be chosen in a way that 
   gives information about  the physical problem whose 
    solution is sought. For the two-body problem, we choose this coordinate by 
    imposing a condition that allows it to be interpreted, near infinity, as the 
    (inverse of the) Newtonian potential. In this way, two quantities 
    that specify the problem -- the separation of the two particles and their mass 
    ratio -- enter the equations from the very beginning. 
    The asymptotic solution (near infinity) is obtained 
    and a natural identification of the Bondi ``news function" in terms of the 
    source parameters is suggested,  leading to an expression for the radiated 
    energy that differs from the standard quadrupole formula but agrees with 
    recent non-linear calculations. When the free function of time describing the
    separation of the two particles is chosen so as to make the 
    new expression agree with the classical result, closed-form  analytic
    expressions are obtained, the resulting metric approaching the 
    Schwarzschild solution with time. As all physical quantities are defined with respect to the flat metric at infinity, the  physical interpretation of this solution depends strongly on how these definitions are extended to the near-zone and, in particular, how the ``time" function in the near-zone is related to Bondi's null  coordinate.

\end{abstract}


\maketitle

\section{Introduction}

 The characteristic initial value formulation \cite{FRIEDcharIV} of the axisymmetric vacuum Einstein
 equations, first proposed by Bondi and co-workers \cite{Bondi62}, has two distinct advantages over the initial value formulation in
 terms of a space-like hypersurface \cite{SPACEinit}: (i) there are no constraint 
 equations to be satisfied
 by the initial data and (ii) the unknown functions can be determined, 
 one at a time,  by performing an hierarchical series of 
quadratures over the initial data and ``known" quantities (including 
functions of integration from previous 
quadratures) \cite{Bondi62,NU,JWlr}. Thus there are no partial differential equations in three 
independent variables to be solved!

This is achieved by writing the equations in terms of coordinates adapted to
 a series of null hypersurfaces $u=const$, one of which is the initial hypersurface
 (assuming that the spacetime, outside a world tube containing the 
 source, can be foliated by such hypersurfaces), and requiring that
all but one coordinate be constant on the null geodesics (generators) in 
these null hypersurfaces. Then the ``main" equations are ordinary differential
 equations that propagate the initial data along these geodesics (``rays'' 
 or bicharacteristics).  The radial coordinate -- the one that varies along 
 the rays -- must be specified by imposing one additional condition. 

There are two standard choices for this additional condition, one  of
 which is always made, resulting in the so-called ``area" or ``affine" radial 
coordinate, respectively. After either choice is made and the equations 
integrated, the dynamics of the two-body problem must be deduced by 
interpreting the properties of the two bodies (masses, separation) in terms of the 
 functions of integration in the chosen coordinates.
  This is a non-trivial matter involving a lot of approximations 
  \cite{Bondi62,GHW}.
  
Now, for a unique solution, additional (boundary) data must be given on a 
timelike (a world tube) or null hypersurface intersecting the initial $u=const$ hypersurface.
Such a  timelike boundary hypersurface is used in the ``Cauchy Characteristic 
Matching" (CCM) approach \cite{JWlr,CCM} and is conventionally defined as having 
constant ``radial coordinate", while in the interior (Cauchy) problem it is 
identified with a coordinate sphere (in a different coordinate system).

In this paper we propose that the choice of  radial coordinate should be made in a way 
that is appropriate to the geometry of the physical problem, i.e., that 
the surfaces of constant radial coordinate must be identified with the level surfaces of a 
physical quantity. In this way, the definition of the radial 
coordinate implicitly defines the physical problem. Applying this idea to
the two-body problem, we impose a condition that 
follows from the requirement that the two-surfaces of constant radial coordinate 
in any null hypersurface share a 
geometric property with the surfaces of constant Newtonian gravitational potential of two point 
particles in Euclidean 3-space. This choice has the advantage that important 
physical parameters, such as the separation\footnote{The word ``separation" does not 
mean ``physical distance". 
For the sense in which physical  terms are used in this paper, see the {\it Remark} at the 
end of  section III.} of the two bodies and their mass 
ratio, appear in the equations from the very beginning. They thus act as ``source" 
terms for this problem. We then integrate the equations, assuming a series 
expansion near infinity and asymptotically flat boundary conditions. By considering the angular dependence of the ``mass" 
parameter of the solution, the arbitrary function of two variables 
generating the 
solution is then naturally identified with the first time derivative 
of the quadrupole moment per unit mass of the source.

Section II  introduces the notation we will be using and section III describes the 
choice of radial coordinate. In section IV we integrate the Einstein equations
 assuming a series expansion of the unknowns near infinity. In section V we 
use physical arguments to determine  the  arbitrary function in the solution, arriving at a result which differs from Bondi's.  We argue that this discrepancy is due to our asymptotic definitions of physical quantities.   In section VI we discuss the validity of our results, 
pointing out that a similar choice for the radial coordinate might be convenient for 
investigating the near field also. Finally, in section VII, we state 
explicitly all assumptions in our approach, compare our derivation to Bondi's and clarify  
the conditions under which our calculations can be interpreted 
as modelling the real physical problem.  

\section{Formulation of the Problem - Notation}

We will restrict our considerations to the the case where the axial Killing vector 
is  hypersurface orthogonal (no particle spin). We will use the symbols 
${u,\: \xi,\: \eta,\: \varphi}$ for the four coordinates. The assumption that the Killing 
vector trajectories (parametrized by $\varphi$) are hypersurface 
orthogonal implies that $g^{\,u\,\varphi}=g^{\,\xi\,\varphi}=g^{
\,\eta\,\varphi}=0$, while the 
requirements that  $u$ is a null coordinate and the curves along which 
only $\xi$ varies are null geodesics imply \cite{Bondi62} that 
$g^{\,u\,u}=g^{\,u\,\eta}=0$ also. Thus there are 5 remaining metric functions 
of ${u,\: \xi,\: \eta}$, denoted by the letters $B,\: V,\:K,\: R,\: U$.
We will write the line element in terms of a canonical  Newman-Penrose 
null frame\footnote{We use the standard NP 
notation \cite{PandR}: the complex null-tetrad basis $\{\el,\: \en,\: \emm, \:\emmB \}$ is 
normalized to $\el \cdot\en = -\emm\cdot\emmB= 1$. We will 
restrict the symbols $\{\el, \:\en, \:\emm, \:\emmB \}$ to denote the 
basis co-vectors (differential forms) while the 
vectors (differential operators) will be denoted by the standard symbols 
$\{D, \:\Delta, \:\delta, \:\deltaB \}$.} as  $ds^{2}=\el\otimes\en+\en\otimes\el-\emm\otimes
\emmB-\emmB \otimes \emm$, where:
\bea
\el =\frac{1}{B}\dx{u},\hspace{30pt}\en 
=\dx{\xi}+\frac{V}{2}\dx{u},\nn \hspace{30pt} 
\emm = \frac{-1}{\rt}\left( \frac{K^{2}}{R}\left(\dx{\eta}-U 
\dx{u}\right)+i\,R\: \dx{\varphi}\right),\hspace{20pt} \nn \\
 \Delta = B \left(\ddx{u}-\frac{V}{2}\ddx{\xi}+U \ddx{\eta}\right),\hspace{20pt}D 
=\ddx{\xi}, \hspace{30pt}\ \delta 
=\frac{1}{\rt}\left(\frac{R}{K^{2}}\ddx{\eta}+\frac{i}{R}\ddx{\varphi} 
\right).\hspace{40pt}\label{NPframe}
\eea
The frame transformations preserving the \el \ direction and keeping 
\emM \ parallel to the Killing vector are a boost in the \el -\en \ plane and 
a null rotation around \el\ parametrized by $S$ and $T$, respectively:
\bea
\el\rightarrow\el/S, \hspace{15pt}\en\rightarrow S[\en+T (\emP) + 
T^{2}\,\el],\hspace{15pt} 
\emm\rightarrow \emm+T\, \el. 
\eea
If one chooses $S=\Xi_{,\,\xi},\: \: T=-R \,
\Xi_{,\,\eta}/(\rt\,K^{2}\,\Xi_{,\,\xi})$ and redefines $B,\: V$ and $U$ appropriately, the new frame is the canonical frame 
that is obtained by a redefinition of the radial coordinate 
$\xi\rightarrow\Xi(u,\,\xi,\,\eta)$. Thus an additional condition must 
be imposed to eliminate this freedom, effectively reducing the number of unknown 
functions of three variables from 5 to 4.
Bondi's standard notation (and choice of radial coordinate, $K=\xi$) is obtained by making the substitutions:  $\xi\rightarrow r,\:\eta\rightarrow -\cos 
\theta,\:B\rightarrow\exp(-2\beta),\: V\rightarrow V_{B}/r, \:
K\rightarrow r,\:U\rightarrow\sin \theta\: U_{B},\:R\rightarrow r\sin \theta \,
\exp(-\gamma)$. 

The reason for using  $\xi,\:\eta$ instead of the more usual
$r,\:x(=\cos\theta)$ is that we want to reserve the latter symbols 
for the radial and angular coordinates of a Euclidean prolate 
spheroidal coordinate system, which we will use to impose a condition 
on the function $K(u,\,\xi,\,\eta)$ in the next section. Here we note 
that, in such a coordinate system (with singularities at $\pm a $ on 
the z-axis), the Schwarzschild solution for a mass $M$ at $z=a$ on 
the symmetry axis is given by the  substitutions: 
\bea
\xi\rightarrow r-a\,x,\hspace{8pt}\eta\rightarrow 
\frac{r\,x-a}{r-a\,x},\hspace{8pt}B\rightarrow 1,\hspace{8pt} V\rightarrow 
1-\frac{2M}{r-a\,x},\nn \\
K\rightarrow r-a\,x,\hspace{8pt}U\rightarrow 0,\hspace{8pt}R\rightarrow 
\sqrt(r^{2}-a^{2})\,\sqrt(1-x^{2}).\label{SCHWprol}
\eea
In the limit $a \rightarrow 0$, we obtain the standard form in spherical 
$r,\:x$ coordinates. What is more, \pref{SCHWprol}
remains a solution if we allow $a$ to be a 
\textit{function}, $a(u)$, as 
it is simply a change of coordinates applied to the Schwarzschild solution 
in ``spherical'' $\xi,\,\eta$ coordinates:
\be
B=1,\hspace{15pt}V=1-\frac{2M}{\xi},\hspace{15pt}K=\xi,\hspace{15pt}U=0,
\hspace{15pt}R=\xi\:\sqrt(1-\eta^{2}).\label{SCHWspher}
\ee

The 2-surfaces of constant $u,\:\xi$ play an important role in the 
characteristic initial value formulation of the Einstein equations 
 \cite{FRIEDcharIV,JWlr,CCM}. In particular, the induced metric on them,
\be
d\sigma^{2}=\frac{K^{4}}{R^{2}}\,\dx{\eta}^{2}+R^{2}\,\dx{\varphi}^{2}, \label{ds2surf}
\ee
is known as part of the initial data. Thus imposing conditions on 
this part of the metric, in a sense, defines the physical problem whose 
evolution is determined by Einstein's equations. Of course, only the $\eta$ 
dependence of $K$ is determined by the initial data, and this can be eliminated 
by a redefinition of the $\eta$ coordinate. This is the approach usually taken 
in imposing the ``spherical area'' radial coordinate condition $K=\xi$. We will instead 
impose a different condition that will allow these 2-surfaces to have different 
shapes depending  on the ratio $a(u)/\xi$, where $a(u)$ is a measure of the 
separation of the two particles.

\section{The geometry of the surfaces of constant Newtonian potential}
Prolate spheroidal coordinates (in Euclidean 3-space) is the 
appropriate coordinate system to use for the two  particle problem. 
The Euclidean line element in such a coordinate system with 
singularities at $\pm a $ on the z-axis is 
\be
ds^{2}_{FLAT}=(r^{2}-a^{2}x^{2})\left(\frac{\dx{r}^{2}}{r^{2}-a^{2}}
+\frac{\dx{x}^{2}}{1-x^{2}}\right)+(r^{2}-a^{2})(1-x^{2})\,\dx{\varphi}^{2}, \label{ds2FLAT}
\ee
and the solution of the Laplace equation for the Newtonian potential 
of two point particles with masses $m_{1}$ at $z=a$ and $m_{2}$ at $z=-a$ is
\be
V_{N}=-\frac{m_{1}}{r-a\,x}-\frac{m_{2}}{r+a\,x}.\label{NEWTpot}
\ee
For the 1-particle problem (say $m_{2}=0$), Bondi's geometrical coordinate choice 
$K=\xi=r-a\,x$, admits several interpretations in terms of 
properties of the surfaces of constant Newtonian potential:
\begin{itemize}
    \item  
$K$ is constant on the surfaces of constant Newtonian potential
    \item 
$2/K^{2}$ is the Gaussian curvature of the surfaces of constant 
Newtonian potential
    \item  $1/K^{4}$ is the flat-space norm of the gradient
    ($\eta^{a\,b}V_{N,a}V_{N,b}$) of the Newtonian potential of a unit mass.
\end{itemize}
It is tempting to impose a condition, following from such an interpretation of $K$, 
on the metric functions in  \pref{NPframe}, 
with $(V-1)/2$ playing the role of the Newtonian 
potential. However, apart from ambiguities in relating the metric 
function $V$ to the ``Newtonian potential'', involving the function $B$, one must remember that
the function $V$ is not a part of the data on the initial $u$ 
hypersurface and, therefore, relating it to $K$ for all $\xi$ (and $u$) may 
unduly restrict the evolution. We will instead impose a condition 
that will determine $K(u,\,\xi,\,\eta)$ without reference to the 
other metric functions.

To do this we write the flat-space metric \pref{ds2FLAT} in terms of 
an orthogonal system of coordinates $\xi,\,\eta,\,\varphi$, where 
the $\xi$ coordinate is the (inverse of the) Newtonian potential 
\pref{NEWTpot} per unit mass:
\be
\frac{m_{1}+m_{2}}{\xi}=\frac{m_{1}}{r-a\,x}+\frac{m_{2}}{r+a\,x}, \hspace{30pt}
(m_{1}+m_{2})\,\eta=m_{1}\,\frac{r\,x-a}{r-a\,x}+m_{2}\,\frac{r\,x+a}{r+a\,x}.\label{ksetDEF}
\ee
Let us denote by $N2(\phi,\,\psi)$ the inner product of the gradients of 
the functions  $\phi,\,\psi$ with respect to the flat metric 
\pref{ds2FLAT}:
\be
N2(\phi,\,\psi)\equiv 
\frac{(r^{2}-a^{2})\,\phi_{,\,r}\,\psi_{,\,r}+(1-x^{2})\,\phi_{,\,x}\,\psi_{,\,x}}{r^{2}-a^{2}x^{2}}.\label{NORMdef}
\ee
Then it is easy to verify that $\xi,\,\eta$ as functions of $r,\,x$ given by 
\pref{ksetDEF} satisfy $N2(\xi,\,\eta)=0$. It is remarkable that orthogonality is maintained when the 
``radial'' and ``angular'' coordinates of the individual 1-particle problems [see 
\pref{SCHWprol}] are combined linearly in this simple way. 
This is undoubtedly due to the appropriateness of 
 prolate spheroidal coordinates for describing this problem. In terms 
 of a parameter $\lambda$  depending on the ratio of the masses, 
 the coordinate transformation \pref{ksetDEF} can be written
 \be
\xi=\frac{r^{2}-a^{2}x^{2}}{r+\lambda\,a\,x},\hspace{25pt}
 \eta=x-\frac{a\,(\lambda\,r+a\,x)(1-x^{2})}{r^{2}-a^{2}x^{2}},\hspace{25pt}
 \mbox{where  }\hspace{20pt}\lambda \equiv \frac{m_{1}-m_{2}}{m_{1}+m_{2}}.\label{KSETdef}
\ee
We note that, when $x \rightarrow\pm 1,\text{ also }\eta \rightarrow \pm 
1$, while  as $r \rightarrow \infty,\: \xi  \rightarrow r \text{ and } \eta  \rightarrow x$.
Thus, for $r\,\gg \,a$, 
the coordinates  $\xi,\,\eta$  behave as prolate spheroidal (or spherical) coordinates $r,\,x$.

In terms of the new coordinates defined in \pref{KSETdef}, the flat 
metric \pref{ds2FLAT} takes the form
\be
ds^{2}_{FLAT}=\frac{\dx{\xi}^{2}}{N2(\xi,\,\xi)}
+\frac{\dx{\eta}^{2}}{N2(\eta,\,\eta)}+(r^{2}-a^{2})(1-x^{2})\,\dx{\varphi}^{2}.\label{ds2FLATkset}\\
\ee
Now, it is easy to verify that 
$N2(\eta,\,\eta)=N2(\xi,\,\xi)(r^{2}-a^{2})(1-x^{2})/\xi^{4}$, so that the metric on the 
$\xi=const$ surfaces in Euclidean 3-space can be put in the form 
\pref{ds2surf} with
\bea
R^{2}=(r^{2}-a^{2})(1-x^{2}),  \hspace{20pt } 
K^{4}=\frac{\xi^{4}}{N2(\xi,\,\xi)}=\frac{\xi^{4}}{1+S},   \text{    where }\label{Kdef}\\
S=\frac{a^2(1-\lambda^{2})(-r^2 + 3\,r^2\,x^2 + 4\,\lambda\,a\,r\,x^3 + {a}^2\, 
x^2+ \lambda^2\,{a}^2\,x^4)}{(r+\lambda\,a\,x)^4}.\label{Sdef}
\eea
The denominator of $S$ (or $\xi$) cannot vanish as the coordinate ranges are 
$r\geq a,\:-1\leq x\leq 1$, and the parameter $\lambda$ is by definition  
less than one in absolute value.

We are now ready to make our choice for the Bondi ``radial'' coordinate  $\xi$ that
is appropriate for the two particle problem: we will require that $K$  
 will be given by \pref{Kdef} \textit{for any 
value of }$u$, i.e., that the parameter $a$ in \pref{Sdef}, 
determining the separation of the two particles, will be allowed to be a function of 
$u$. 

To be used in Einstein's equations $K$ must, of course, be given 
as a function of $u,\,\xi,\,\eta$ by inverting the coordinate 
transformation \pref{KSETdef}. This cannot be done in closed form, as 
it depends on the roots of a 5th order polynomial in $\xi,\,\eta$.  
However, it is easy to convert the series expansion of  $K$ for $r 
\rightarrow \infty$ to one for $\xi \rightarrow \infty$, making use of 
the limits noted after \pref{KSETdef}. This is most easily done by replacing, 
in the series expansion of $S$ with respect to $r$, the symbols
$r,\, x \text{ by } \xi,\,\eta$, respectively, one order at a time 
and expanding again:
\bea
  S \simeq \frac{(1 - \lambda^2)\, (-1 + 3\,x^2)\,{a(u)}^2}{r^2} 
-\frac{4\,\lambda\,( 1 - \lambda^2) \,x\,( -1 + 2\,x^2 ) \,{a(u)}^3}{r^3}+ 
{O(\frac{1}{r})}^4,\nn\\
  S-\frac{(1 - \lambda^2)\, (-1 + 
3\,\eta^2)\,{a(u)}^2}{\xi^2}  \simeq -\frac{4\, \lambda\, \left(1 - {\lambda}^2 
\right)\,x\, \left(-3 + 5\, x^2 \right)\,{a(u)}^3}{r^3} + 
{O(\frac{1}{r})}^4,\nn \\
  S -\frac{(1 - \lambda^2)\, (-1 + 
  3\,\eta^2)\,{a(u)}^2}{\xi^2} 
+\frac{4\, \lambda\, \left(1 - {\lambda}^2 \right)\, 
\eta\, \left(-3 + 5\, \eta^2 \right)\, {a(u)}^3}{\xi^3}\simeq 
{O(\frac{1}{r})}^4.
\eea
Substituting in \pref{Kdef} we obtain, to this order,
\be
  K \simeq \xi-\frac{(1 - \lambda^2)\, (-1 + 3\,\eta^2)\,{a(u)}^2}{4\:\xi} 
+\frac{\lambda\, \left(1 - {\lambda}^2 \right)\, \eta\, \left(-3 + 5\, \eta^2 \right)\, 
{a(u)}^3}{\xi^2}+{O(\frac{1}{\xi})}^3. \hspace{10pt}\label{Kseries}
\ee
In this way we can compute the series expansion of $S$, and therefore 
of $K$, as a function of $u,\,\xi,\,\eta$ to any desired order.  We note that the angular dependence of the first two ``correction terms" in $K$ are  proportional to the Legendre polynomials $P_{2},\, P_{3}$ and therefore vanish when integrated over angles. Thus we can  interpret our coordinate $\xi$, alternatively, as an approximate ``luminosity distance": the area of the surfaces of constant $u$ and $\xi$ equals $4\,\pi \,\xi^2$ to a good approximation. To find out how good this approximation is we must carry the expansion to 5 more orders! We then find (see \pref{ds2surf}):
\be
\int d\mathcal{S}= \int K^2 d\eta\,d \varphi = 4\,\pi \,\xi^2\left(1-\frac{4}{35}{(1 - \lambda^2)}^2(1 -6\, \lambda^2)\frac{a(u)^6}{\xi^6}+{O(\frac{1}{\xi})}^7\right).
\ee
Thus  our radial coordinate $\xi$  is also a ``luminosity distance" to a very good approximation.

It should be pointed out that the reflection symmetry, $\eta 
\rightarrow -\eta$, assumed in many investigations \cite{GHW,LBGSW}, 
  holds here 
only if, at the same time, $\lambda \rightarrow -\lambda$. This is to 
be expected, as the interchange of the positive and negative 
$z$-directions will not give the same problem unless the two 
particles are also interchanged.

Finally, we compute the flat-space quadrupole moment of the two particles 
relative to their center of mass. In terms of cylindrical 
coordinates $\rho,\,z,\,\varphi$
with origin at the center of mass, the mass density distribution is
\be
\mu = 
\frac{\delta(\rho)}{2\,\pi\,\rho}\{m_{1}\delta[z-(1-\lambda)\,a(u)]
+m_{2}\delta[z+(1+\lambda)\,a(u)]\}.
\ee
Then the quadrupole moment tensor can be written
\be
D^{\,a\,b}=\int \mu\,(3\,x^{a}x^{b}-x^{2}\delta^{\,a\,b})\,d\,V
=Q(u)\,(3\,\delta^{a}_{\:z}\delta^{b}_{\:z}-\delta^{\,a\,b}),  \label{QMtensor}
\ee
where
 \bea
Q(u)&= \int 
\{m_{1}\delta[z-(1-\lambda)\,a(u)] +m_{2}\delta[z+(1+\lambda)\,a(u)]\}\,z^{2}\,dz \nn \\
&=(m_{1}+m_{2})(1-\lambda^{2})\,a(u)^2,\hspace{40pt} \label{Qdef} 
\eea
and we have used the definition of $\lambda$  to write  $m_{1},\,m_{2}$ in terms 
of $m_{1}+m_{2}$ and $\lambda$.

\emph{Remark:} All statements regarding physical quantities (particle separation, mass density,
quadrupole moment, etc.) in this paper refer to an unphysical background 
Minkowski space, which is  the limit as $\xi \rightarrow \infty$ of 
the sought solution of Einstein's equations. 
In the physical, curved, spacetime they are to be understood only as
names referring to the combination of variables in their definition. 
Thus, for example, ``particle separation" means $2\,a(u)$ and not physical distance between 
the two particles.

\section{Series solution of the vacuum Einstein equations}

Knowing $K$ and assuming a formal series expansion for the function $R$, 
valid near $\xi \rightarrow \infty$, where $R$ approaches the 
flat-space limit \pref{SCHWspher}, we can carry out the well-known hierarchical 
series of $\xi$ integrations to obtain the other metric functions. Thus, we 
assume that as $\xi \rightarrow\infty$, $R$ has the series expansion
\be
R\simeq \xi \,
\sqrt(1-\eta^2)
 \left(1+\frac{c_{1}(u,\,\eta)}{\xi}+\frac{c_{2}(u,\,\eta)}{\xi^2}
+\frac{c_{3}(u,\,\eta)}{\xi^3}+{O(\frac{1}{\xi})}^4\right).\label{Rseries}
\ee
Then the Ricci component $\Phi_{0\,0}$ in the frame defined in 
\pref{NPframe} can be solved for $B_{\,\xi}/B$, 
giving
\be
\frac{B_{\,\xi}}{B}=-\frac{K_{\,\xi}}{K}+2\,\frac{R_{\,\xi}}{R}-\frac{K_{\,\xi\,\xi}}{K_{\,\xi}}
-\frac{K\,{R_{\,\xi}}^2}{R^2\,K_{\,\xi}}.
\ee
Integrating the series expansion of the rhs when $K,\:R$ are given by equations
\pref{Kseries}, \pref{Rseries}, and choosing the function of $u,\,\eta$ of integration 
to satisfy the boundary condition \pref{SCHWspher} at infinity, we 
find\footnote{We only give the leading two terms in each equation. All 
calculations were carried out using \textit{Mathematica} and the author's ``RGTC'' 
package (\url{http://www.inp.demokritos.gr/~sbonano/RGTC/}), keeping terms 
to one order higher than in \pref{Kseries}, \pref{Rseries}, i.e., to four 
orders beyond flat space.
The complete expressions can be  found in the \textit{Mathematica} notebook 
``NewRadialCoord.nb'', available at 
\url{http://www.inp.demokritos.gr/~sbonano/TwoBody/NewRadialCoord.nb}.}
\be
B\simeq 1 + \frac{(1 - 3\, {\eta}^2 )\,
      (1 - {\lambda}^2)\, {a(u)}^2 +2\, {c_{1}(u,\, \eta)}^2}
{4\, {\xi}^2}+\ldots
\ee
We next compute the series expansion of the Weyl tensor component 
$\Psi_{0}$, obtaining\footnote{In the 
frame defined in \pref{NPframe}, \emph{all} NP quantities 
(spin-coefficients, Ricci and Weyl tensor components) are real -- for 
a proof see the Appendix in \cite{SB03}.}
\be
\Psi_{0}\simeq -\frac{(1 - 3\, {\eta}^2 )\,
      (1 - {\lambda}^2)\, {a(u)}^2 + 2\, {c_{1}(u,\, \eta)}^2-4\,c_{2}(u, \,\eta)}
{2\, {\xi}^4}+\ldots 
\ee
Now, the condition for asymptotic flatness (absence of incoming radiation 
at infinity) is that \cite{NU}
$\Psi_{0}\simeq {O(\frac{1}{\xi})}^5$, so that we must require that
\be
c_{2}(u, \eta)=\frac{(1 - 3\, {\eta}^2 )\,
      (1 - {\lambda}^2)\, {a(u)}^2 + 2\, {c_{1}(u,\, \eta)}^2}{4}.\label{c2SOL}
\ee
Next, the Ricci component $\Phi_{0\,1}$ can be written
\be
{[\frac{B\,K^{\,6}\,U_{\,\xi}}{R^{\,2}}]}_{\xi}= 
K^2\,[\frac{B_{\,\xi}\,B_{\,\eta}}{B^2}-\frac{B_{\,\xi\,\eta}}{B}
+2\,\frac{R_{\,\xi}\,R_{\,\eta}}{R^2}+2\,\frac{R_{\,\xi\,\eta}}{R}]
+ 2\,K\,K_{\,\xi}[\frac{B_{\,\eta}}{B}-2\,\frac{R_{\,\eta}}{R}],
\ee
so that, with the series expansions of $K,\,R,\,B$ known, $U$ can be 
obtained. We find
\bea
U \simeq  \frac{(1-\eta^2)\,c_{1,\,\eta}(u, \,\eta)-2\,\eta \,c_{1}(u,\, 
\eta)}{\xi^2}\hspace{100pt} \nn \\
+\frac{(1-\eta^2)[4\,\,c_{1}(u, \,
\eta)\,c_{1,\,\eta}(u, \,\eta)-x_{0}(u,\, \eta)]-8\,\eta \,c_{1}(u, \,
\eta)^2}{3\,\xi^3}+\ldots,
\eea
where $x_{\,0}(u,\, \eta)$ is an arbitrary function of integration. The second such function, 
to be added to $U$, must be set equal to zero for asymptotic flatness.

Proceeding in the same way, equation $\Phi_{1\,1}+3\,\Lambda=0$ can be solved for
$(V\,K\,K_{\,\xi})_{\,\xi}$ in terms of known quantities, and, with $V$ known, equation  
$\Phi_{0\,2}=0$ gives $(K\,R_{,\,u}/R)_{\,\xi}$. Denoting by $y_{\,0}(u,\, 
\eta),\:z_{\,0}(u,\, \eta)$ the functions of integration,  $V$ and 
$R_{,\,u}$ are given by
\bea
&V \simeq 1 + \frac{y_{\,0}(u,\, \eta)}{\xi}+\ldots   \label{Vseries}\\
 &R_{,\,u} \simeq \sqrt(1-\eta^2)[z_{\,0}(u,\, \eta)
+\frac{2\,c_{1}(u, \,\eta)\,z_{\,0}(u,\, \eta)+(1 - 3\, {\eta}^2 )\,
(1 - {\lambda}^2)\, a(u)\,\dot{a}(u)}{2\,\xi}]+\ldots 
\eea
Finally the requirement that this agrees with the $u$ derivative of \pref{Rseries} 
determines the  $u$ derivatives of the coefficients $c_{i}(u, \,\eta)$ 
(except for $c_{2}(u, \,\eta)$ which is given 
by \pref{c2SOL}):
\be
c_{1,\,u}(u, \,\eta)= z_{\,0}(u,\, \eta),\hspace{25pt}
c_{3,\,u}(u, \,\eta)=\ldots \label{EVOLeq}
\ee

This completes the integration of the so-called ``main'' equations. 
Of the remaining equations, $\Phi_{1\,1}=0$ is satisfied identically, while the 
vanishing of  $\Phi_{1\,2},\:\Phi_{2\,2}$ impose the following two 
conditions -- conservation laws -- on
the three functions of integration  $x_{\,0}(u,\, 
\eta),\:y_{\,0}(u,\, \eta)$ and $z_{\,0}(u,\, \eta)=c_{1,\,u}(u, \,\eta)$:
\bea
&x_{0, \, u}=- y_{0, \, \eta} + 
c_{1}\,c_{1, \, u \, \eta} - 3\,c_{1, \, u}\,c_{1, \, \eta} 
+ 3\, \eta\,(1 - \lambda^2)\, a(u)\, \dot{a}(u) \nn \\
&y_{0, \, u} = 2\, c_{1, \, u}^2 -2\, c_{1, \, u}+ 
(1 - \eta^2)\,c_{1, \, u \, \eta \, \eta} -4\, \eta\,c_{1, \, u\, \eta} \label{EMcons}\\
&+(1 - 3\, \eta^2)(1 - \lambda^2)[{\dot{a}(u)}^2 + a(u)\,\ddot{a}(u)].\nn
\eea

The series solution obtained in this section reduces to the one with 
Bondi's coordinate choice $K=\xi$ in the limit $\lambda^2\rightarrow 1$. In 
fact it can be obtained as a coordinate transformation from the Bondi 
solution if one redefines the $c_{i}$ to be given in terms of the 
corresponding $c_{i\,B}$ (= coefficients in the series expansion of $R$ 
in  powers $K$) by the expressions obtained when $K$ is replaced by 
its series expansion 
\pref{Kseries}. One then finds that $c_{1}=c_{1\,B}$ and $c_{2}$ is given 
by \pref{c2SOL}. To make the conservation laws \pref{EMcons} take the Bondi form, 
one must also redefine the functions of integration $x_{\,0}(u,\, 
\eta),\:y_{\,0}(u,\, \eta)$ as follows:
\bea
&x_{\,0}(u,\, \eta)\rightarrow x_{\,0\,B}(u,\, 
\eta)+\frac{9}{2}\,\eta\,(1 - \lambda^2)\,{a(u)}^2,\label{x0y0Bondi} \\
&y_{\,0}(u,\, \eta)\rightarrow 
y_{\,0\,B}(u,\, \eta)+(1 -3\, \eta^2)(1 - \lambda^2)\,a(u)\,\dot{a}(u).\nn
\eea
Despite the fact that, formally, the solution obtained here is but a 
coordinate transformation of the Bondi solution,  the 
explicit appearance of the ``source'' terms involving $a(u)$ and $\lambda$ 
in the conservation laws \pref{EMcons} makes the choice of the arbitrary 
function $c_{1}$ describing the 
two-body problem and, consequently, the physical interpretation of the 
solution particularly simple and transparent.

\section{Particular solution describing the two-body problem}
The solution obtained in the previous section is a ``general'' 
solution of the equations in that it depends on the arbitrary 
function $c_{1}(u, \,\eta)$. To obtain the solution for a 
particular problem, an appropriate choice for this function must be 
made. With the coordinates chosen to fit the two-body problem, it 
is reasonable to expect that the required arbitrary function will have a simple 
form. First, non-singular behavior on the axis ($\eta=\pm1$) requires that 
$c_{1}(u, \,\eta)=q(u, \,\eta)\,(1-\eta^2)$ for some $q(u, \,\eta)$ 
that is well behaved at $\eta=\pm1$. 
Making this substitution in the second conservation equation 
\pref{EMcons} and replacing $y_{\,0}(u,\, \eta)$ by $-2M(u,\, 
\eta)$ (Bondi's  ``mass aspect"  definition -- see equation \pref{Vseries}), we obtain
\bea
&2\, M_{, \, u} =- 2\, {(1 - {\eta}^2 )}^2\, {q_{, \, u}}^2 \nn \\
&-(1 - 3\, {\eta}^2 )\{(1 - {\lambda}^2 )\, [{\dot{a}(u)}^2 + a(u)\,\ddot{a}(u)] - 
4\, q_{, \, u}\}  \label{MassLoss} \\
&- (1 - {\eta}^2 )[(1 - {\eta}^2 ) q_{, \,u\, \eta\, \eta}
 - 8\, \eta\, q_{, \, u\, \eta}].\nn
\eea
Now, the rhs of this equation describes the energy loss of the 
system. 
On physical grounds, it must be negative definite and have the angular dependence that is appropriate to the one-dimensional motion of the two particles. This angular dependence, 
being independent of $\xi$, can be identified with the angular 
distribution of the flow of energy 
at infinity obtained in the linearized theory using the Landau-Lifshitz pseudotensor
(see \cite{LL} equation 110.15)
\be
 \frac{d\,I}{d\,\Omega}\sim
[\frac{1}{4}{(\dddot{D}_{a\,b}n^{a}n^{b})}^2+\frac{1}{2}\dddot{D}_{a\,b}\dddot{D}^{a\,b}
-\dddot{D}^{a}_{\:\,b}n^{b}\dddot{D}_{a\,c}n^{c}],
\ee
where $\dddot{D}_{a\,b}$ is the third time derivative of the quadrupole
moment tensor\footnote{It is consistent to use the flat-space definition \pref{QMtensor} for 
the quadrupole moment tensor here, as the L-L pseudotensor is defined in terms of the  linear solution to the field equations, which is determined by a flat-space integral over the sources. Besides, only the angular dependence of $D_{a\,b}$ is used in the present argument, not the exact form of $Q(u)$.} \pref{QMtensor} and  $n^{a}$
are the Cartesian components of the unit vector in the direction of 
propagation, which in our coordinates equal $(\sqrt(1-\eta^2)\cos 
\varphi,\:\sqrt(1-\eta^2)\sin \varphi,\:\eta)$ at infinity. We find
\bea
\frac{d\,I}{d\,\Omega}\sim 
[\frac{1}{4}{(3\,\eta^2-1)}^2+\frac{1}{2}(6)-(3\,\eta^2+1)]\hspace{20pt} \nn \\
=[{\left(\frac{3\,\eta^2-1}{2}\right)}^2+1-(3\,\eta^2-1)]=\frac{9}{4}{(\eta^2-1)}^2.
\eea
Thus the
angular dependence of the mass-loss equation \pref{MassLoss} will be 
$\sim {(1 - {\eta}^2)}^2$, as appropriate for this system, if we choose 
\bea
&q\,(u,\, \eta)=\frac{1}{4}(1 - {\lambda}^2 
)\,a(u)\,\dot{a}(u),\hspace{10pt}\text{so that}\label{q0SOL}\\
&c_{1}(u, \,\eta)=\frac{1}{4}(1 - {\lambda}^2 )\,a(u)\,\dot{a}(u)\,(1-\eta^2)
=\frac{1}{8}\dot{\cQ}\,(1-\eta^2)\label{c1SOL},
\eea
where  $\cQ \equiv (1 - {\lambda}^2 )\,{a(u)}^2$  is the quadrupole 
moment per unit mass -- see \pref{Qdef}. The same result follows from 
the  requirement that the rhs of \pref{MassLoss} be negative for all 
$u,\,\eta$, so that energy flows \emph{out} of the system in all directions 
and at all times: \pref{q0SOL} is the unique solution that is regular on the axis and 
makes the linear (and ``source" terms) on the rhs of \pref{MassLoss} vanish.

With  this $c_{1}(u, \,\eta)$, the first conservation equation can be 
integrated giving
\be
x_{\,0}(u,\, \eta)=\eta\,(1 - {\lambda}^2 
)\,{a(u)}^2\,[\frac{3}{2} +\frac{(1-\eta^2)(1 - {\lambda}^2 )\,{\dot{a}(u)}^2}{8}]
+2\int M_{, \, \eta}\,d\,u, \label{x0int}
\ee
where  $M(u,\,\eta)$ is determined by the equation
\be
  M_{, \, u} =-{(1 - {\eta}^2 )}^2\, {\left(\frac{(1 - {\lambda}^2 
)\,[{\dot{a}(u)}^2 + a(u)\,\ddot{a}(u)]}{4}\right)}^2
=-{(1 - {\eta}^2 )}^2\, {\left(\frac{\ddot{\cQ}}{8}\right)}^2 \label{Mint}
\ee
once the function $a(u)$ is known.

Finally, using $\cQ$,
the quadrupole moment per unit mass of the system, the leading terms 
in the components of the 
Weyl tensor in the frame defined in 
\pref{NPframe}, with $c_{1}(u, \,\eta)$ given  by \pref{c1SOL}, are:
\bea
& \Psi_{0}=\frac{6}{{\xi}^5}\,[c_{3}(u,\, 
\eta)+\lambda\,\eta\,(3-5\,\eta^2)\,\cQ\,a(u)]+\ldots, \nn \\
 &\Psi_{1}=\frac{\sqrt(1-\eta^2)}{\sqrt 2\:{\xi}^4}¥\left(\int M_{, \, 
\eta}\,d\,u-\frac{3}{2}\,\eta [ \cQ
+\frac{(1-\eta^2)\,{\dot{\cQ}}^2}{16} ]
\right)+\ldots,\nn \\
&\Psi_{2}=\frac{-1}{{\xi}^3}\left(M+\frac{(1-3\,\eta^2)\dot{\cQ}}{4}+
\frac{(1-\eta^2)^2\,\dot{\cQ}\,\ddot{\cQ}}{64}\right)+\ldots, \label{PSIsol} \\
& \Psi_{3}=\frac{\eta\,\sqrt(1-\eta^2)\,\ddot{\cQ}}{2\sqrt 2\:\xi^2}+\ldots,  \hspace{15pt} 
 \Psi_{4}=\frac{(1-\eta^2)\,\dddot{\cQ}}{8\:\xi}+\ldots \nn
 \eea
Using the evolution equations \pref{EVOLeq}, it can be shown that $\Psi_{0}$  is proportional to $(1-\eta^2)$.

We note that equation \pref{Mint}, giving the 
radiated energy directly in terms of the parameters $\lambda$ and 
$a(u)$ describing the source, 
is an \emph{exact} result following from a particular, physically motivated, 
choice of the arbitrary function $c_{1}(u, \,\eta)$ (or $q(u, \,\eta)$).
And this choice of $q(u, \,\eta)$, which eliminates all linear terms on the rhs of 
the energy-conservation equation \pref{MassLoss}, is made possible by the 
existence of the extra term introduced by our choice of the function $K$, 
which fortuitously  has the correct angular dependence $(1-3\,\eta^2)$.
The final expression for the mass loss \pref{Mint}, relating the radiated energy to the square of the 
second ``time" derivative of the ``quadrupole moment per unit mass'' 
differs from the standard quadrupole formula (third time derivative 
of the quadrupole moment) obtained in the linearized theory. However, it closely resembles the results of approximate non-linear 
calculations, where the radiated energy is found to be proportional to the square of 
an \emph{integral} with respect to $u$ of $\dddot{Q}/M$ weighted 
by $\exp(2 u/M)$ \cite{NonLinRad1,NonLinRad2}, 
where $M$ is a large constant ``background'' mass.  
Thus our choice of $c_{1}(u, \,\eta)$, equation \pref{c1SOL}, 
implying that $M_{, \, u} \sim {\ddot{\cQ}}^2$ in the fully 
non-linear case, is in remarkably good agreement with these results. The discrepancy with the classical result must be attributed to the different definitions of  ``time" and ``quadrupole moment per unit mass''  (see the \emph{Remark} at the end of section III). And the following considerations reinforce this conclusion.

Bondi's approximate \emph{linear} calculation for the news function, 
which agrees with the conclusions of linearized theory 
\cite{LL} and post-Newtonian calculations \cite{FutSchutz, WinQRF}, 
gives   $c_{1}(u, \,\eta)=-\frac{1}{2}\ddot{Q}(u)\,(1-\eta^2)$ 
(in our notation Bondi's $c$ equals $- c_{1}$).
If we want to reconcile the two results we must require that, as a 
consequence of the equations of motion\footnote{The equations of 
motion are used in the derivation of the result in linearized  or 
post-Newtonian theory 
\cite{LL,FutSchutz}, and, implicitly, in Bondi's derivation.},
the quadrupole moment satisfies the equation $\ddot{Q}(u)=-\frac{1}{4}\,\dot{Q}(u)/(m_{1}+m_{2})$ so that,
\bea
Q(u)=Q_{0}\,\exp[-\frac{u-u_{0}}{4\,(m_{1}+m_{2})}], \hspace{10pt} \text{and 
therefore,} \nn \\
a(u)=a_{0}\,\exp[-\frac{u-u_{0}}{8\,(m_{1}+m_{2})}].\hspace{70pt}\label{EqMotGuess}
\eea
Of course, $a(u)$ should be determined by the equations of 
motion following from the vanishing of the divergence of  the energy momentum
tensor of the two particles. 
But this requires knowledge of the field in the vicinity of the 
particles, which is well beyond the scope of the approximate calculations near 
null infinity carried out here. 

Nevertheless, we point out that, with $a(u)$ given by
 \pref{EqMotGuess}, \textit{all} remaining $u$ integrations 
(see equations \pref{EVOLeq}, \pref{x0int}, \pref{Mint}) can be
evaluated analytically in terms of elementary functions (exponentials); 
and choosing the functions (of $\eta$) of integration to vanish 
(except for the final mass), the resulting
solution will smoothly approach the Schwarzschild solution as 
$a(u)\rightarrow 0$ exponentially with 
$u \rightarrow \infty$.  Now, if we assume that the Minkowskian definitions of  ``separation'', ``velocity'' and ``time"  used here approximate adequately the corresponding physical 
quantities, this solution has the unphysical feature 
that the two particles approach each other at a diminishing rate (both 
$a(u)$ and $\dot{a}(u)$ decrease with time): the collision seems to end with hardly a 
whimper rather than a bang! However, the physical interpretation of ``velocity" depends on the definition of ``time": near the source the time function, 
which can be taken to equal $u+\xi$ at infinity, is expected to have a logarithmic 
singularity, as in the Schwarzschild case \cite{Israel66, SPW95}. A time function of the
 form $t=u+\xi-(1+k)a(u)+8\,(m_{1}+m_{2})\, \log(\xi/a_{0})$ (where $k$=const) evaluated on the curve $\xi=a(u)$ gives $t=u_{0}-k\, a(u)$, so that $d\,a/d\,t=-1/k$. Thus, with a proper definition of time in the near-zone, the rate of approach implied by  \pref{EqMotGuess}  might be physically acceptable. It seems too speculative to discuss further this possibility. But this example shows that one cannot simply identify $u$-derivatives at infinity with time-derivatives in the near-zone, and that the mass-loss formula \pref{Mint} may be consistent with the standard quadrupole formula!
 
In any case, this particular solution, being an exact analytic
solution (in the form of a series expansion which can be continued to any 
order) of the Einstein equations, {\it including the evolution equations}, 
which approaches the Schwarzschild solution as $u\rightarrow\infty$,
can be useful in testing the accuracy of numerical codes. For this reason,
explicit  expressions for the metric functions and the Weyl tensor 
components, together with the verification that the metric is 
Ricci-flat to the appropriate order and becomes Schwarzschild as $u \rightarrow \infty$, 
are included in the \textit{Mathematica} notebook 
``NewRadialCoord.nb'' mentioned in footnote 3.

\section{Discussion}
The freedom in the choice of coordinates inherent in general relativity 
is invariably used to simplify the equations. This  is due to  
the complexity of the equations. However, a coordinate choice that best
simplifies the equations may not describe the physical problem in the 
most natural way. For example, in the static, axially symmetric problem, 
Weyl's canonical $\rho,\:z$ coordinates require that a physically 
spherical source be described as a linear distribution of mass.

In the Bondi formulation of the axisymmetric vacuum equations, the 
essential simplification comes from the use of a null coordinate $u$ 
and angular coordinates $\eta, \:\varphi$ which are \textit{constant} 
on the null rays. The  choice of parametrization on these rays does 
not simplify  the mathematical problem any further. In this paper we 
propose that it should be made in a way that gives information about 
the particular physical problem whose solution is sought. For the asymptotic solution near 
infinity of the two-body problem, imposing a condition that follows 
from  interpreting the surfaces on which $u,\:\xi$ are 
constant as surfaces of constant Newtonian potential, allows us to 
relate  the arbitrary function of two variables generating the solution to 
two parameters describing the source, $\lambda$ and $a(u)$, 
by a symmetry argument \textit{without making any approximations}. 
This should be compared to Bondi's  approximate and, in his own words, 
``distinctly crude''\footnote{See  comments after equation (91) in  
\cite{Bondi62}. An outline of Bondi's derivation is given in section VII.} 
derivation of the corresponding result. 

Despite its simplicity and directness, our derivation of the form of 
$c_{1}$ is subject to a serious objection, as
the result seems to depend on a choice of coordinates. How can such an 
arbitrary choice  lead to a physically meaningful result? There are two 
answers to this criticism:
(i) Once the arbitrary function in the solution ($q(u,\,\eta)$) has been related to the arbitrary function ($\cQ(u)$) determining the coordinate transformation \pref{Kseries}, the latter is no longer arbitrary.  The same result could have been obtained with Bondi's choice of 
radial coordinate $\xi_{B}=K$, had one chosen the arbitrary functions 
$x_{\,0}(u,\, \eta),\:y_{\,0}(u,\, \eta)$ to include extra source terms as in 
equation \pref{x0y0Bondi}, and then transformed the radial coordinate 
from $K$ to $\xi$ to eliminate the extra term in the ``mass aspect"\footnote{This extra term in the ``mass aspect"\ does not change the ``Bondi Mass", as its integral over angles vanishes. In fact, in obtaining  his formula for $c_{1B}$, Bondi also obtains a first-order expression for $M_{B}$ that has precisely the form $\dot{f}(u)(1-3\cos^2\theta)$, as in \pref{x0y0Bondi}: in going from eq. 88 to eq. 89 of reference \cite{Bondi62}, Bondi has set his function $p(u)$ equal to $2\,Q_{00}$ (plus a  constant $M0$)!} (this is discussed further in the next section). In this sense, the leading term in the coordinate transformation \pref{Kseries} is determined by the solution.
(ii) Our choice of radial gauge, following from the identification, near infinity, of 
the two-surfaces of constant $u,\,\xi$ with the surfaces of constant 
Newtonian potential, is not an arbitrary choice of coordinates but is 
closely related to the physics: the surfaces of constant 
$\xi$ are the wave fronts of the radiation emitted from the system, 
carrying information about its properties; 
and identifying their constant time sections ($t=u+\xi$ near infinity) with the surfaces of constant Newtonian potential gives a  description that matches the dynamics 
 more accurately than the use of either  spherical or  prolate-spheroidal 
\cite{GHW} surfaces (any change in the dynamics at infinity must be manifested through changes in the Newtonian potential). In fact, the dynamics can be better approximated 
if one allows the parameters $m_{1},\,m_{2}$ to depend on $u$ to 
reflect the relativistic velocity dependence of mass. All $\xi$ integrations 
remain unchanged and only equations involving $u$ derivatives of the 
unknown functions will
acquire extra terms, leading to more complicated evolution equations 
\pref{EVOLeq} and conservation laws \pref{EMcons}. Indeed, the 
energy loss equation \pref{Mint}, expressed in terms of $\ddot{\cQ}$, remains 
unchanged even when the parameter $\lambda$ in $\cQ$ is allowed to depend on $u$.

This leads to the suggestion that, in the inner problem also, 
a coordinate condition based on a property of
appropriately defined constant-Newtonian-potential 
(or constant-Newtonian-energy) surfaces  should be  used, 
as these surfaces have the topology of the ``pair of pants'' 
picture of the horizon \cite{JWlr}:  near the source, 
the set of points with constant Newtonian 
potential consists of two disconnected subsets, one around each 
particle. This can best be seen if, near $\xi=0$, one switches to the $r,\: x$ 
coordinates used in the definition of $\xi,\: \eta$ \pref{KSETdef}. Of 
course, near the source the null coordinate $u$ will not be well 
behaved and a different time coordinate must be used, while care must 
be exercised in defining the ``Newtonian potential''.

\section{Summary and conclusions}

This paper addresses the mathematical problem of solving Einstein's
equations with the symmetries (one hypersurface-orthogonal
Killing vector with closed spacelike orbits) and boundary conditions
(asymptotically flat, no incoming radiation) that are appropriate for 
the two-particle collision problem. All calculations are carried out 
near infinity, and physical quantities are defined with respect to the 
limiting Minkowski space (see the  {\it Remark}  at the end of section III).
The Bondi formulation of this problem is used 
as it allows free initial data and a
sequential integration of the equations. Both the solution and the
coordinate system are only assumed to be valid outside some 
sufficiently large time-like world tube. 

A particular, explicit, solution to this mathematical problem depends on
two arbitrary choices: (a) the ``radial" coordinate condition -- the
coordinate expression for the determinant of the metric on the two-surfaces of
constant $u,\,\xi$ (the function $K(u,\, \xi,\, \eta)$). (b) the free
function $c_{1}(u, \,\eta)$. Mathematically, these choices are completely
arbitrary (modulo boundary conditions).

The basic idea of this paper is to try to
``guess" the appropriate form for these arbitrary functions based on the
``expected" properties of the solution describing the two-particles; properties  
which can be calculated in the region where the solution is valid, 
i.e., outside the large world tube. Thus,  (a) the
function $K(u,\, \xi,\, \eta)$ is chosen by the requirement that, on any surface of 
constant $u$ and $\xi$, it is identical to the corresponding function
in the metric of  the surfaces of constant Newtonian potential of two point
particles in Euclidean 3-space. Even though the expression for $K$ in
Euclidean space is valid everywhere, we only use the asymptotic form of
$K$, valid for, say, $\xi > 10\,a(u)$.  (b) The function $c_{1}(u, \,\eta)$ (or
$q(u, \,\eta)$) is chosen by the requirement that, at infinity, the angular
dependence of the emitted radiation is that appropriate to quadrupole
radiation. The same $q(u, \,\eta)$ satisfies the physical requirement 
that a non-negative amount of energy is radiated in any direction and at 
any time.

The physical significance of allowing $K$ to have a term $f(u,\,\eta)/\xi$ is that 
it changes the ``mass-aspect". This is most easily seen by making the change of coordinates
$r \rightarrow  \tilde{r}+p(u,\,\theta)/ \tilde{r}$ to the Schwarzschild metric in 
null-spherical coordinates, $ds^2=(1-2M/r)du^2+2\,du \,dr-r^2d\Omega^2$. Then, to first 
order in $1/ \tilde{r}$, the metric acquires an extra term $2\, du\, d\theta\,  
p_{\theta}/ \tilde{r}$ but the ``mass aspect" (=coefficient of $1/ \tilde{r}$ in $g_{u\,u}$) 
is also changed to $M-p_{u}$ due to a contribution from the $2\,du \,dr$ term in the metric. 
This is the basic reason for the existence of the extra term on the rhs of the mass-loss 
equation \pref{MassLoss} for the metric defined by \pref{NPframe}, which allows the  
linear terms to vanish, leaving $\dot{M}=-{\dot{c_{1}}}^2$,  when one chooses 
$K=\xi-f(u)(3\,\eta^2-1)/\xi+\ldots$ and $c_{1}=\frac{1}{2}\dot{f}(u)(1-\eta^2)$. 
One could then make the  ad-hoc choice $f(u)=\dot{\tilde{Q}}(u)$ and claim that one has 
obtained the standard quadrupole formula as an exact result of the non-linear equations!  
But this ad-hoc choice lacks any physical justification for calling this $\tilde{Q}$ the 
``quadrupole moment". Our interpretation of $\xi$ in terms of the Newtonian potential at infinity 
gives both the correct angular dependence to the $1/\xi$ term in $K$ \emph{and provides a physical 
interpretation of the coefficient} $f(u)$, namely $Q(u)/(4M)$ (with $Q$, admittedly, defined 
in a flat background). Of course, the question of uniqueness remains: could a different physical interpretation of the ``radial coordinate" $\xi$ lead to a function $K$ that (i) agrees with Bondi's $K=\xi$ both at infinity and for the 1-particle limits ($\lambda \rightarrow \pm 1,\mbox{ or }a(u)\rightarrow 0$), (ii) the limit $\mathop {\lim }\limits_{\xi \to \infty } \xi(K-\xi)$ exists and has the form $f(u)(3\,\eta^2-1)$  but with a different physical meaning of $f(u)$?

It is appropriate, at this point, to recall how Bondi obtains his relation between $c_{1}$ and 
the quadrupole moment. He begins with the static, axially symmetric vacuum metric in Weyl 
coordinates
\bea
e^{2\,\psi}dt^2-e^{-2\,\psi}\,[e^{2\,\sigma}(d\rho^2+dz^2)+\rho^2 d\varphi^2],\nn
\eea
in which the function $\psi$ satisfies the flat-space Laplace equation. He then
\begin{enumerate}
\item{Takes for $\psi$ the solution (in spherical coordinates $\rho=R\sin \Theta,\;\;z=R\cos\Theta$)
\bea
\psi=-\frac{m}{R}-\frac{D\cos\Theta}{R^2}-\frac{(Q+1/3\,m^3)(3\cos^2\Theta-1)}{2\,R^3},\nn
\eea
$m,\,D$ and $Q$ being called the {``mass", ``dipole", ``quadrupole moment"} of the source. 
The term $1/3\,m^3$ is needed to make $Q=0$ for the Schwarzschild solution, as it is obtained from a linear 
mass distribution in these coordinates.}
\item{Finds the coordinate transformation to Bondi coordinates $(t,\,R,\,\Theta) \rightarrow 
(u,\,r,\,\theta)$ as a series approximation for large $r$ (or $R$).}
\item{By comparing coefficients of the transformed (to Bondi coordinates) Weyl static solution 
with the series expansion of the metric \pref{NPframe}, 
he finds expressions for  the unknown functions $c_{3},\,x_{0},\,y_{0}$ 
(equivalent to his $C,\,N,\,M$) in terms of 
$m,\,D,\,Q$. The function $c_{1}$ is also obtained as a coordinate-dependent expression.}
\item{Next he supposes that, for slowly varying fields, he can use the same expressions for these functions
allowing the constants ${m,\,D,\,Q}$ to be functions of $u$. And, for weak fields, he solves 
the \emph{linearized} form of the equations \pref{EVOLeq},  \pref{EMcons} determining the $u$ derivative of 
$c_{3},\,x_{0},\,y_{0}$ (expressed now in terms of ${m,\,D,\,Q}$) for  the $u$-dependence of 
$D,\,m,\,c_{1}$, respectively. In this way $Q$, instead of $c_{1}$, becomes the free function.}
\end{enumerate}

It will be observed that the last step involves two serious approximations. Bondi is alluding to 
this in calling his derivation ``distinctly crude". In particular, what conclusions would have 
he obtained had he allowed the constants ${m,\,D,\,Q}$ to be time-dependent \emph{before making 
the coordinate transformation}, and, perhaps, having found first a time-dependent first-order correction to the static solution?  This crucial step (allowing the constants to be functions of $u$) is necessary in order to generalize the static Newtonian potential to a dynamic one, but it is also the weakest point in the derivation: when $\psi$ is $u$-dependent, the Weyl metric is not a solution (to first order in $\dot{m},\,\dot{D},\,\dot{Q}$) of the equations!

In our approach, we make contact with a $u$-dependent Newtonian potential via our interpretation of $\xi$,  which we are free to do as it is a coordinate condition, \emph{not an approximation for slowly varying fields}. Moreover, we do not need to integrate the linear approximations of \pref{EVOLeq},  \pref{EMcons}  to relate the free function $c_{1}$ to the properties of the source. 
We obtain  $c_{1}$ directly by a physical argument: it is the unique, non-singular solution that  makes the linear terms on the rhs of \pref{MassLoss} vanish, resulting in an \emph{exact} expression for the mass loss 
 ($\dot{M}=-{\dot{c_{1}}}^2$) that has both the correct angular dependence and is negative definite.
If we accept these arguments as physically reasonable, then we are obliged to conclude that the radiation depends on the ``second derivative of the quadrupole moment per unit mass"
 --  a result which seems to disagree with the standard quadrupole formula. We must remember, however, that our physical quantities (and those of references \cite{NonLinRad1,NonLinRad2}) are defined at infinity and the ``dots" in our formula cannot be translated directly to ``time derivatives"  in the near-zone. On the other hand, the quadrupole formula has been proved \cite{QFproof} only in a \emph{linear} (in the metric tensor) approximation to the field equations. 

 The entire solution depends on an
arbitrary function of one variable, $a(u)$, that is qualitatively related
to the particle separation. This function is determined by the
equation of motion which, however, cannot be evaluated at infinity.  

In conclusion, we wish to stress that the solution obtained in this 
paper (with the choice \pref{c1SOL} for $c_{1}$) is a formal asymptotic approximation (for large $\xi$) 
to an exact solution of the full non-linear Einstein equations that has all the required properties for describing the two particle collision problem. The solution can be continued to any 
desired order in the expansion and appears to converge for $\xi > 
\,a(u)$ in the sense that no large numerical coefficients appear. 
However, it is safer to assume that it is only valid for, say,  $\xi > 
10\,a(u)$. The interpretation of the solution, which was 
also the motivation for making the above choices for the free 
functions $K$ and $c_{1}$, avoids mathematically unjustified steps in Bondi's interpretation; however, it involves physical quantities which
 are defined only in a background Minkowski space -- the limiting form of 
the solution as  $\xi\rightarrow\infty$. Using this solution to calculate the 
outer boundary data for the inner problem in a CCM approach \cite{JWlr,CCM},
the physical definition of the function $a(u)$  and of ``time" in the near-zone can, in principle, be obtained.  

\emph{Note added in proof:} It can be shown that the form of $K$ given in \pref{Kseries} is valid for any mass distribution. Specifically, if we use spherical coordinates ${r,x}$ and define $\xi(r,x)$ by the series expansion  (Newtonian potential of a unit-total-mass distribution)
\bea
\frac{1}{\xi}=\frac{1}{r}+\frac{M_{1}(u)\,x}{r^2}+\frac{M_{2}(u)\,(3\,x^2-1)}{2\,r^3}+\frac{M_{3}(u)\,x\,(5\,x^2-3)}{2\,r^4}+\ldots \nn
\eea
and $\eta(r,x)$ by the requirement that $\nabla{\xi}\cdot  \nabla{\eta}=0$, then, proceeding as in Section III,  we find that
\bea
K=\xi+\frac{k_{1}(u)\,(3\,\eta^2-1)}{\xi}+\frac{k_{2}(u)\,\eta\,(5\,\eta^2-3)}{\xi^2}+\ldots, \nn
\eea
where $k_{1}(u)=[{M_{1}(u)}^2-M_{2}(u)]/4$ and $k_{2}(u)=-{M_{1}(u)}^3+\frac{3}{2}M_{1}(u)M_{2}(u)-\frac{1}{2}M_{3}(u)$. Thus, when the dipole-moment vanishes, $k_{1}(u)=-M_{2}(u)/4=-Q/(4\,M)$ for any mass distribution.

\begin{acknowledgments}
I want to thank my colleague and friend G. K. Savvidy for his enthusiastic support and 
encouragement and for several stimulating discussions. I also thank Professor S. Persides 
for valuable comments and suggestions. Finally, I must thank the referees for their criticisms 
which have helped me improve the presentation and interpretation of the results.
\end{acknowledgments}

\end{document}